\documentclass[fleqn,usenatbib]{mnras}

\usepackage{newtxtext,newtxmath}

\usepackage[T1]{fontenc}
\usepackage{ae,aecompl}

\usepackage{graphicx}	
\usepackage{amsmath}	

\def\pmj0818{PM~J08186$-$3110}
\title[Magnetic DZ white dwarf]{The nearby magnetic cool DZ white dwarf PM~J08186$-$3110}

\author[A. Kawka, et al.]{Adela Kawka$^{1}$\thanks{Contact e-mail: \href{mailto:adela.kawka@curtin.edu.au}{adela.kawka@curtin.edu.au}}, St\'ephane Vennes$^{2}$,
Nicole F. Allard$^{3,4}$,  T. Leininger$^{5}$ and 
\newauthor
F.~X. Gad\'ea$^{5}$
\\
$^1$ International Centre for Radio Astronomy Research - Curtin University, GPO Box U1987, Perth, WA 6845, Australia\\
$^2$ Mathematical Sciences Institute, The Australian National University, Canberra, ACT 0200, Australia\\
$^3$ GEPI, Observatoire de Paris, Universit\'e PSL, CNRS, UMR 8111, 61 avenue de l'Observatoire, F-75014 Paris, France\\
$^4$ Sorbonne Universit\'e, CNRS, UMR 7095, Institut d'Astrophysique de Paris, 98bis boulevard Arago, F-75014 Paris, France\\
$^5$ 4 Laboratoire de Chimie et Physique Quantiques, UMR 5626, Universit{\'e} de Toulouse (UPS) and CNRS, 118 route de Narbonne,\\
F-31400 Toulouse, France}

\date{Accepted XXX. Received YYY; in original form ZZZ}

\pubyear{2020}

\begin{document}
\label{firstpage}
\pagerange{\pageref{firstpage}--\pageref{lastpage}}
\maketitle

\begin{abstract}
We present an analysis of photometric, spectroscopic and spectropolarimetric data of the nearby, cool, magnetic DZ white dwarf \pmj0818. High dispersion spectra show the presence of Zeeman splitted spectral lines due to the presence of a surface average magnetic field of 92~kG. The strong magnesium and calcium lines show extended wings shaped by interactions with neutral helium in a dense, cool helium-rich atmosphere. We found that the abundance of 
heavy elements varied between spectra taken ten years apart but we could not establish a time-scale for these variations; such variations may be linked to surface abundance variations in the magnetized atmosphere. Finally, we show that volume limited
samples reveal that about 40\% of DZ white dwarfs with effective temperatures below 7000~K are magnetic.
\end{abstract}




\section{Introduction}

Magnetic fields are detected among all spectral classes of white dwarfs. The incidence of magnetism
was reported to be significantly higher in white dwarfs polluted with heavy elements compared to the incidence of magnetism
in the general white dwarf population \citep{kaw2014,hol2015,kaw2019}. In polluted hydrogen-rich (DAZ) 
white dwarfs, the incidence of magnetism is $\approx 50$~\% for white dwarfs with effective temperatures
$T_{\rm eff} < 6000$~K and all of these have magnetic field strengths below 1~MG \citep{kaw2014,kaw2019}. 
Only two magnetic DAZ white dwarfs are known with $T_{\rm eff} > 6000$~K, NLTT~53908 
\citep[$T_{\rm eff} = 6250$~K,][]{kaw2014} and WD~2105$-$820 \citep[$T_{\rm eff} = 9890$~K,][]{lan2012,gen2018}. The incidence of magnetism in pure-hydrogen DA white dwarfs, i.e., those apparently devoid of heavy elements, may also be age dependent with a higher prevalence at cooler temperatures, $T_{\rm eff} \sim 5000$~K \citep{rol2015}. A study of the local sample \citep{gia2012} does indicate a peak incidence of 20\% among cool DA white dwarfs \citep{kaw2019}. However, the polluted, magnetic DA white dwarfs (DAZH) show a much narrower field spread ($B < 1$~MG) than their unpolluted counterparts (DAH) and they appear to form a distinct population. It is not known whether the magnetic fields in these two populations have a common origin.

In the case of polluted helium-rich white dwarfs (DZ),\footnote{In this paper, the spectroscopic label ``DZ'' simply refers to polluted,
helium-rich white dwarfs, and the label ``DAZ'' refers to polluted, hydrogen-rich white dwarfs.  However, we note that ultra cool ($<<5000$~K), polluted, hydrogen-rich white dwarfs with vanishing Balmer lines (not encountered in the present study) may be assimilated to the DZ white dwarfs in the same fashion that ultra cool DA white dwarfs maybe assimilated to the featureless DC white dwarfs.} the incidence of magnetism is about $22$~\% for white dwarfs
with $T_{\rm eff} < 8000$~K and magnetic field strengths ranging from a few hundred kG up to about 30~MG \citep{hol2015,hol2017}. SDSS~J1242+0829 is the only known magnetic DZ with $T_{\rm eff} > 8000$~K \citep{cou2019}. Continuum polarization measurements may reveal the presence of a strong magnetic field
in unpolluted DC white dwarfs. The detection of
continuum polarization in three DC white dwarfs that lie within 20~pc of the Sun \citep{bag2020} suggests that the incidence of magnetism in DC white dwarfs is also high, but with much stronger magnetic fields ($\approx 100$~MG) than in their polluted (DZ) counterparts.

Two scenarios have been proposed to explain the higher incidence of magnetism in cool, polluted white dwarfs. In the first scenario an old white dwarf encounters and collides with a gaseous planet giving rise to
differential rotation in the white dwarf envelope. This differential rotation would then generate a relatively weak magnetic field
\citep{far2011,kaw2019}. The chance of such a collision increases with time and \citet{far2011} estimated
that such a stellar encounter has a 50\% probability of occurring every 0.5~Gyr. In the second scenario proposed
the magnetic field in DAZ and DZ white dwarfs is generated in the core of the parent giant stars \citep{kis2015,hol2017}.
Here, a planet or a low mass stellar companion could generate a dynamo between the radiative and convective
envelopes during the shell burning phases. This dynamo would create the magnetic field which would remain 
hidden below the non-magnetic envelope for about 1 to 2 Gyr. In either scenario, this magnetic field was created
by the engulfment of a planet, it is likely that other planets or asteroids may have survived and over time
they would move toward the white dwarf and become the source of pollution in the white dwarf atmosphere.

The carbon-rich, hot DQ white dwarfs ($T_{\rm eff} \gtrsim 18\,000$~K) form another class of objects showing a significantly higher incidence of magnetism. These stars have an incidence of at least 70\% \citep{duf2013} and they are also
fast rotators, with periods ranging from a few minutes to a couple of days \citep{dun2015,wil2016}. These rare objects also have an average mass that is higher than that of the general white dwarf population. The combined properties suggest that these stars have formed in double degenerate mergers \citep{dun2015,kaw2020} rather than following
the engulfment of a planet as suggested in the case of DZ white dwarfs. \citet{cou2019}
have extended this class of massive DQs to lower temperatures ($10\,000 \la T_{\rm eff} \la 16\,000$~K) and showed
that these stars have a higher carbon abundance than DQ white dwarfs with normal masses (M $\approx 0.6$~M$_\odot$).

\pmj0818\ was identified as a white dwarf candidate by \citet{lep2005} and \citet{sub2005} due to its
colour and large proper motion. Follow-up spectroscopy revealed it to be a cool DZ white dwarf \citep{sub2008} and 
\citet{bag2019} reported \pmj0818\ to have a weak magnetic field based on spectropolarimetric measurements. Finally, a parallax measurement of $52.30\pm0.71$~mas obtained by \citet{sub2017} places \pmj0818 within 20~pc of the Sun.

In this paper, we present a photometric, spectroscopic and spectropolarimetric analysis of \pmj0818. 
Observations of \pmj0818\ are presented in Section 2. Our analysis of the atmosphere and magnetic field
strength is presented in Section 3 and in Section 4 we revisit the incidence of magnetism in DZ white dwarfs. 
Finally, we present our conclusions in Section 5.

\section{Observations}

We first observed \pmj0818\ with the FORS1 spectrograph attached to the Very Large Telescope (VLT) of the 
European Southern Observatory (ESO) at Paranal on UT 2007 October 22. We used the 600 lines mm$^{-1}$ grism 
centred at 4650 \AA. The slit width was set to 1 arcsec  providing a resolution of $\sim 6$ \AA. The observations
were conducted in the spectropolarimetric mode where we obtained an exposure of 480~s with the retarder plate rotated to
$-45^\circ$, immediately followed by another exposure of 480~s with the retarder plate rotated to $+45^\circ$.

These FORS1 spectra revealed \pmj0818\ to be a cool DZ with a weak magnetic field. We also obtained two 1200 s
observations of \pmj0818\ with the R-C spectrograph attached to the 4~m telescope at Cerro Tololo Inter-American 
Observatory (CTIO) on UT 2008 February 23. We used the KPGL2 (316 lines/mm) grating with the WG360 order blocking filter.
We set the slit width to 1.5 arcsec providing a resolution of $\sim 8$ \AA. This spectrum showed the asymmetric \ion{Mg}{i} optical line triplet characteristic of cool DZ white dwarfs and appeared as a lower quality replica of the FORS1 spectrum.

Consequently, we obtained a higher resolution spectrum with the X-shooter spectrograph \citep{ver2011} attached 
to the VLT at ESO. The spectrum was obtained on UT 2017 October 23 with the slit width set to 0.5, 0.9 and 0.6 arcsec 
with total exposure times of 2975~s, 3040~s and 3060~s for the UVB, VIS and NIR arms, respectively. This set up 
provided a resolving power of 9700, 8900 and 5600 for the UVB, VIS, and NIR arms, respectively. We removed telluric
absorption bands using \textsc{molecfit} recipes \citep{kau2015}.

We collected photometric optical measurements from the SkyMapper survey \citep{onk2019} and infrared 
photometric measurements from the Two degree All Sky Survey \citep[2MASS;][]{skr2006} and the Wide-field
Infrared Survey Explorer \citep[WISE;][]{cut2014}. We also obtained astrometric and photometric measurements
from Gaia \citep{gai2018}. Table~\ref{tbl_astrometry} lists available astrometric and photometric data.

\pmj0818\ was observed with the {\it Transiting Exoplanet Survey Satellite} \citep[$TESS$][]{ric2015} in 
Sector 7 from 2019 January 1 to February 2. The spectral range of the $TESS$ bandpass is from $6000$ to $10\,000$ \AA. 
Since \pmj0818\ (TIC 147018085) was a target of interest, 2 min cadence images were available. We used 
the pre-search data conditioning simple aperture photometry. 
    
\begin{table}
 \caption{Astrometry and photometry of \pmj0818.}
 \label{tbl_astrometry}
 \begin{tabular}{cccccc}
  \hline
  \multicolumn{2}{l}{Parameter} & \multicolumn{3}{c}{Measurement} & Ref. \\
  \hline
  \multicolumn{2}{l}{RA (J2000)} & \multicolumn{3}{c}{08$^{\rm h}$18$^{\rm m}$40\fs26} & 1 \\
  \multicolumn{2}{l}{Dec (J2000)} & \multicolumn{3}{c}{-31\degr 10\arcmin 20\farcs3} & 1 \\
  \multicolumn{2}{l}{$\mu_\sigma \cos{\delta}$ (\arcsec yr$^{-1}$)}& \multicolumn{3}{c}{$0.23772\pm0.00004$} & 1 \\
  \multicolumn{2}{l}{$\mu_\delta$ (\arcsec yr$^{-1}$)} & \multicolumn{3}{c}{$-0.78598\pm0.00005$} & 1 \\
  \multicolumn{2}{l}{$\pi$ (mas)} & \multicolumn{3}{c}{$51.65\pm0.03$} & 1 \\
\hline
\multicolumn{6}{c}{Photometry} \\
\hline
Band & Measurement & Ref. & Band & Measurement & Ref. \\
\hline
  $G$   & 15.408$\pm$0.001 & 1 & $i$ & 15.432$\pm$0.005 & 2 \\
  $b_p$ & 15.639$\pm$0.003 & 1 & $z$ & 15.515$\pm$0.006 & 2 \\
  $r_p$ & 15.072$\pm$0.003 & 1 & $J$ & 14.916$\pm$0.036 & 3 \\
  $u$ & 16.659$\pm$0.034 & 2 & $H$ & 14.728$\pm$0.073 & 3 \\
  $v$ & 16.827$\pm$0.016 & 2 & $K$ & 14.829$\pm$0.122 & 3 \\
  $g$ & 15.531$\pm$0.006 & 2 & $W1$ & 14.680$\pm$0.036 & 4 \\
  $r$ & 15.376$\pm$0.017 & 2 & $W2$ & 15.014$\pm$0.070 & 4\\
  \hline
 \end{tabular}\\
References: (1) \citet{gai2018}; (2) \citet{onk2019}; (3) \citet{skr2006}; (4) \citet{cut2014}
\end{table}

\section{Analysis}

New model atmospheres appropriate for the analysis of low-dispersion (FORS1, R.-C. spec.) and high-dispersion X-shooter spectra have been computed with a mixed H/He composition including a selection of trace elements (C, N, O, Na, Mg, Al, Si, Ca, Cr, Mn, Fe, Ni). The models are in a joint radiative and convective equilibrium adopting the mixing-length convection theory. Although the presence of a magnetic field may inhibit convective motion \citep{tre2015}, \citet{bed2017}, \citet{kaw2019} found that deep convection zones cannot be suppressed in cool white dwarfs such as \pmj0818.

\subsection{Models}

\begin{figure*}
    \centering
    \includegraphics[viewport=25 175 555 680,clip,width=1.9\columnwidth]{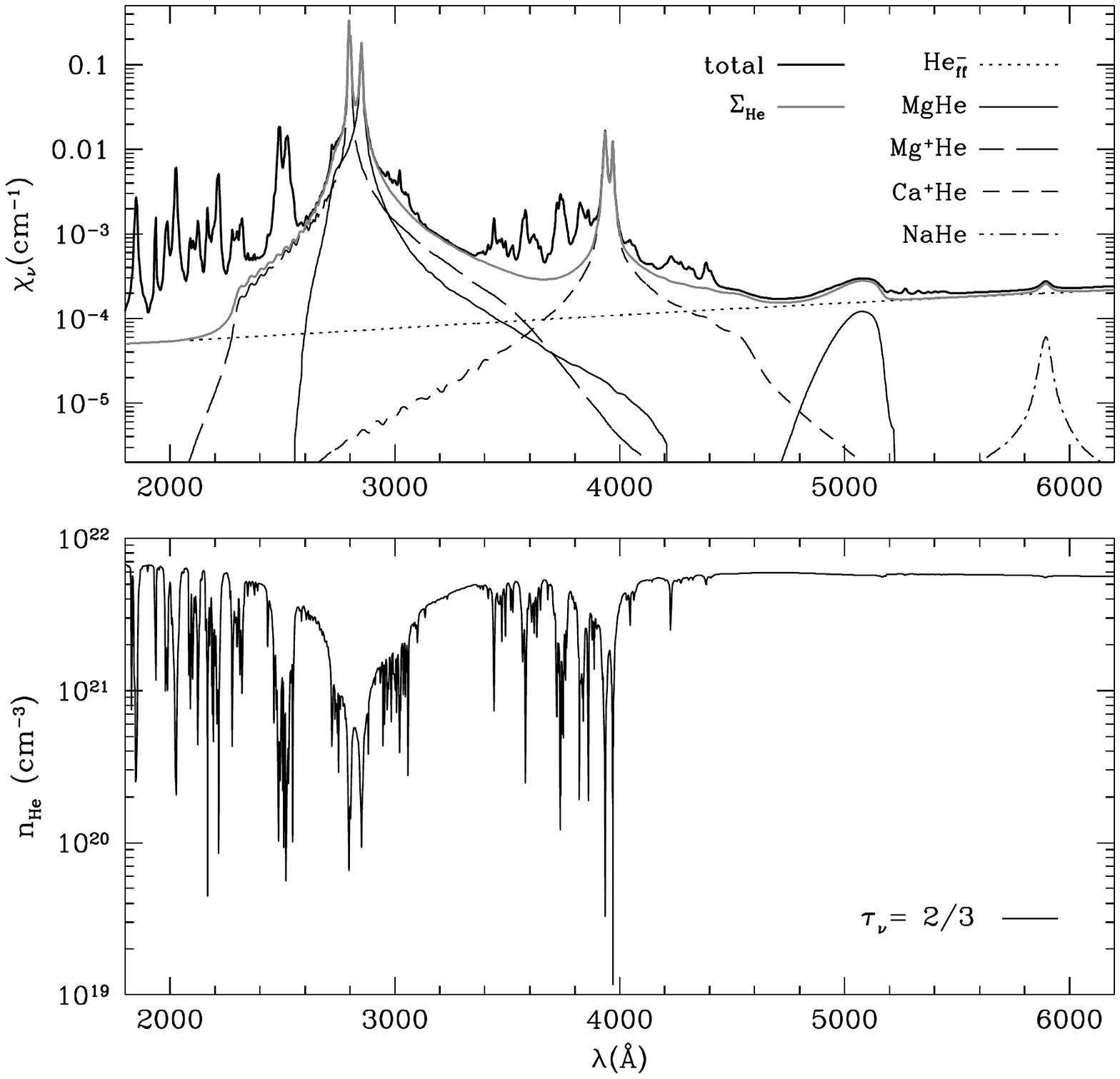}
    \caption{(Top panel) dominant opacities at $\tau_R=0.1$, $n({\rm He})=3\times10^{21}\,{\rm cm}^{-3}$ in a model at $T_{\rm eff}=6\,250$~K and $\log{g}=8.25$. (Bottom panel) neutral helium density at monochromatic optical depth $\tau_\nu=2/3$ in the same model shown in the upper panel.}
    \label{fig_opacity}
\end{figure*}

The model atmospheres are converged to a tolerance of 0.1\% in the depth-dependant integrated flux. The input parameters are the effective temperature ($T_{\rm eff}$), surface gravity ($\log\,g$), and the chemical composition expressed as the number density relative to helium. The total radiative opacity includes dominant contributions displayed in Fig.~\ref{fig_opacity} (top panel) for a representative model. Line profiles of most spectral lines are described by Voigt functions with the width of the Lorentzian component dominated by neutral helium collisions. 
However, cool helium-rich white dwarfs require a specific treatment for line broadening, owing to the high helium densities that are
involved in their atmospheres (bottom of Fig.~\ref{fig_opacity}). New theoretical profiles were very accurately determined in a unified
theory of collisional line profiles valid at very high densities using ab initio calculations of potential energy and transition dipole 
moments. Beyond the symmetrical Lorentzian core at low density, profiles are asymmetrical (see, e.g., for the \ion{Mg}{i} triplet, 
\citep{all2016b} and top of Fig.~\ref{fig_opacity}). These calculations are available for dominant spectral lines (see below). For 
other lines we employ hydrogen line collision parameters from \citet{bar2000} and scaled for helium collisions. 

Fig.~\ref{fig_opacity} (bottom panel) also shows the neutral helium density at a monochromatic optical depth $\tau_\nu=2/3$, which corresponds to a photon escape probability of 50\%, across the optical spectrum for the same model. For example, the \ion{Mg}{i}$\lambda 5178$ triplet and \ion{Na}{i}$\lambda5892$ doublet form at a density $n_{\rm He}\approx 5\times10^{21}$~cm$^{-3}$, while the strong CaH\&K doublet forms at a range of density from $10^{19}$ to $5\times10^{21}$~cm$^{-3}$. 

Details of the line profile calculations that include interactions with neutral helium atoms in dense atmospheres are available for Na-He, Mg-He, Mg$^+$-He, Ca-He, and Ca$^+$-He interactions. The optical \ion{Mg}{i}-He 5167, 5173, 5183\AA\ lines show strong asymmetric blue-shifted wing due to the Mg-He interaction \citep{all2016b}. We tabulated theoretical line profiles at 6\,000~K and helium density at $\log{(n_{\rm HeI}/{\rm cm^{-3}})}=19.0$ to 22.5 in steps of 0.1 dex. The tables encompass the range of neutral helium density found in the atmosphere of the cool DZ white dwarf \pmj0818. A single, representative temperature was adopted for the line forming region. 

The ultraviolet resonance lines \ion{Mg}{ii} 2896, 2803 \AA\ (P$_{1/2}$, P$_{3/2}$) and \ion{Mg}{i} 2852 \AA\ are shaped by Mg$^+$-He and Mg-He interactions, respectively \citep{all2016a, all2018}. The line wings extending toward longer wavelengths are clearly detected in the UVB/X-shooter spectrum. We tabulated the \ion{Mg}{ii} and the \ion{Mg}{i} line cross-sections at neutral helium densities from $\log{(n_{\rm HeI}/{\rm cm}^{-3})}=19.0$ to 22.0 in steps of 0.2~dex, and at 6\,000~K. Lorentzian profiles are adopted at lower densities. 

The \ion{Ca}{i} 4226\AA\ resonance line is shaped by Ca-He interactions \citep{blo2019}. We tabulated the line cross-sections from $\log{(n_{\rm HeI}/{\rm cm}^{-3})} = 20.3$ to 23.0 in steps of 0.1~dex, and at 6\,000~K. Lorentzian profiles were adopted at lower densities with a full-width at half-maximum adjusted to match the Ca-He profile at $\log{(n_{\rm HeI}/{\rm cm}^{-3})} = 20.3$.

The resonance \ion{Ca}{ii} H\&K shaped by Ca$^+$He interactions \citep{all2014} were tabulated at $\log{(n_{\rm HeI}/{\rm cm}^{-3})} = 19.0$, 19.3, 19.6, 19.9, 20.0, 20.3, 20.6, 20.9, 21.0, 21.3, 21.6, 21.9, 22.0, 22.3, and 22.6, covering the range of neutral helium density encountered in the atmosphere of \pmj0818 (Fig.~\ref{fig_opacity}).

Finally, the resonance \ion{Na}{i} is shaped by Na-He interactions \citep{all2013}. We tabulated the line cross-sections at $\log{(n_{\rm HeI}/{\rm cm}^{-3})}=19.0$, to 22.0 in steps of 0.5 dex at a temperature of 6\,000~K.

Other lines currently modelled with Lorentzian profiles should eventually be modelled with line profiles including quasi-molecular interactions, e.g., Fe-He, Ni-He, and others.

\begin{figure}
    \centering
    \includegraphics[viewport=0 170 560 680,clip,width=0.9\columnwidth]{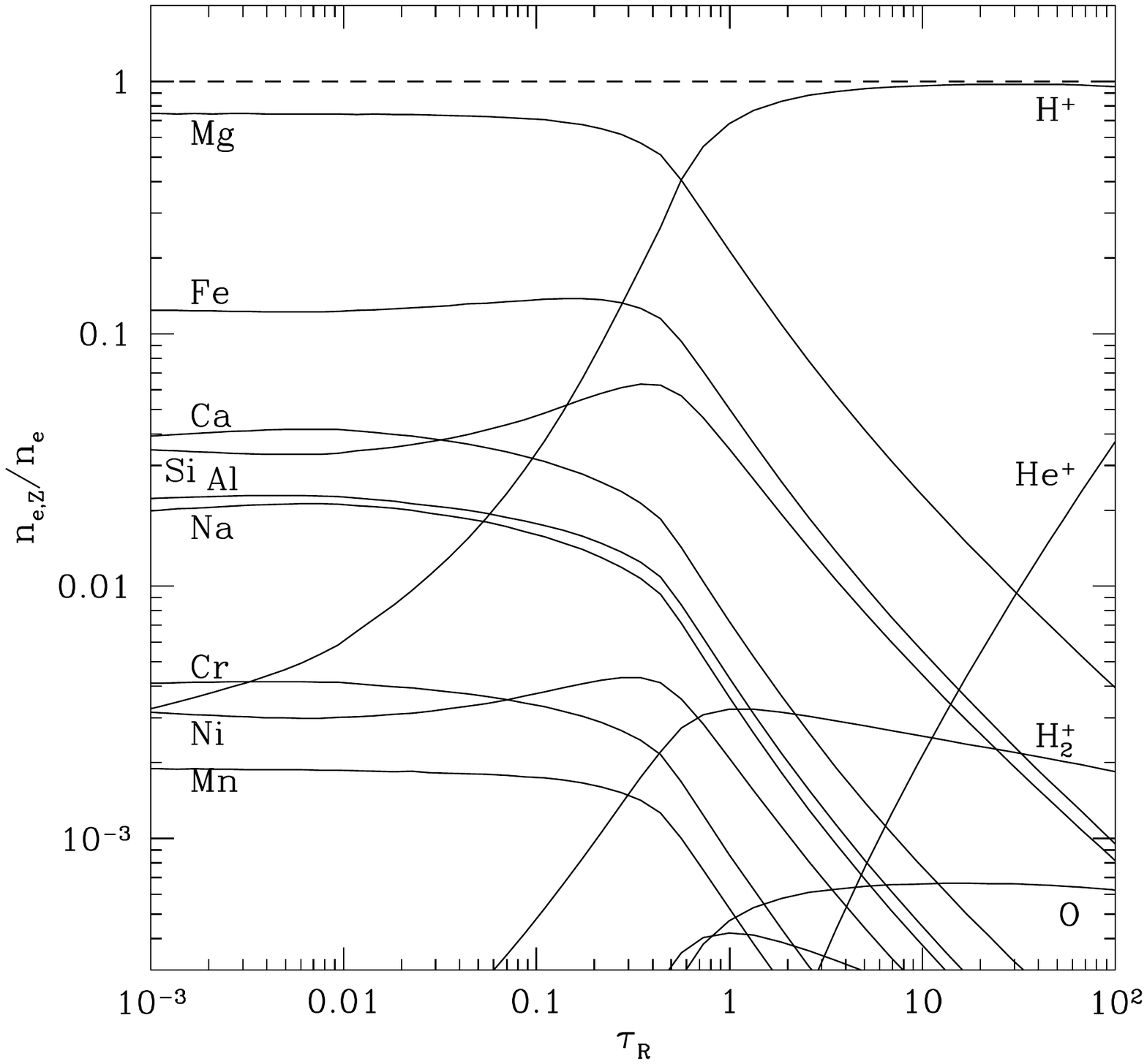}
    \caption{Electron donors expressed as a fraction of the total electron density in a reference model at $T_{\rm eff}=6\,250$~K and $\log{g}=8.25$ and abundances from Table~\ref{tbl_param}.}
    \label{fig_electron}
\end{figure}

The negative helium ion opacity (He$^-$) dominates the optical red to infrared spectrum (Fig.~\ref{fig_opacity}). The main contributors to the electron density are magnesium and hydrogen (Fig.~\ref{fig_electron}). Lowering of the ionization potential in the high density helium environment contributes to an increase in the electron density. We adopted a simple prescription for the ionization potential of C, Na, Mg, Ca, and Fe as a function of temperature and density described in \citet{blo2019b}.

Individual abundances of light and heavy elements are either measured directly or, in the absence of a detection, they are scaled on iron using abundance ratios measured in chondrite meteorites \citep{lod2009}. Lacking clear detections, \citet{cou2019} scaled the abundance of iron ($-7.8$) and other elements on the measured calcium abundance ($-9.0$) using the chondrite meteorites abundance ratios. The presence of hydrogen may be felt indirectly from its large contribution to the electron density and, consequently, the He$^-$ opacity \citep[see ][]{cou2019}. The broad spectral range covered in our spectra should allow us to measure the effect of hydrogen on the continuum shape. \citet{sub2017} fixed the hydrogen abundance at $\log{\rm H/He}=-5$ while \citet{cou2019} fixed the abundance at the detection limit of $\log{\rm H/He}=-3.17$. Other sources of electrons in outer atmospheric layers include elements with low first ionization potential.

A weak field of 92kG was applied to the spectral synthesis (see Section 3.3). At the spectral resolution offered by X-shooter, this effect is apparent in the \ion{Ca}{ii} IR triplet.

\subsection{Stellar parameters}

We first constrained the likely effective temperature and surface gravity with the Gaia distance measurement and the total flux emitted
by \pmj0818.
We integrated the flux from the ultraviolet to the infrared. In the ultraviolet and visual we used the UVB and VIS X-shooter spectra scaled to the SkyMapper photometric
measurements. We extended the flux to the infrared with a model spectrum that is scaled to $2MASS$ and $WISE$ photometric measurements. We then 
calculated the radius using the Gaia distance measurement for a range of effective temperatures. Finally, we used the mass-radius 
relations for hydrogen-deficient white dwarfs of \citet{cam2017} to determine the corresponding mass and surface gravity. Table~\ref{tbl_const} lists the parameters that were constrained by the Gaia distance measurement.

\begin{figure*}
    \centering
    \includegraphics[viewport=23 150 585 695,clip,width=0.9\columnwidth]{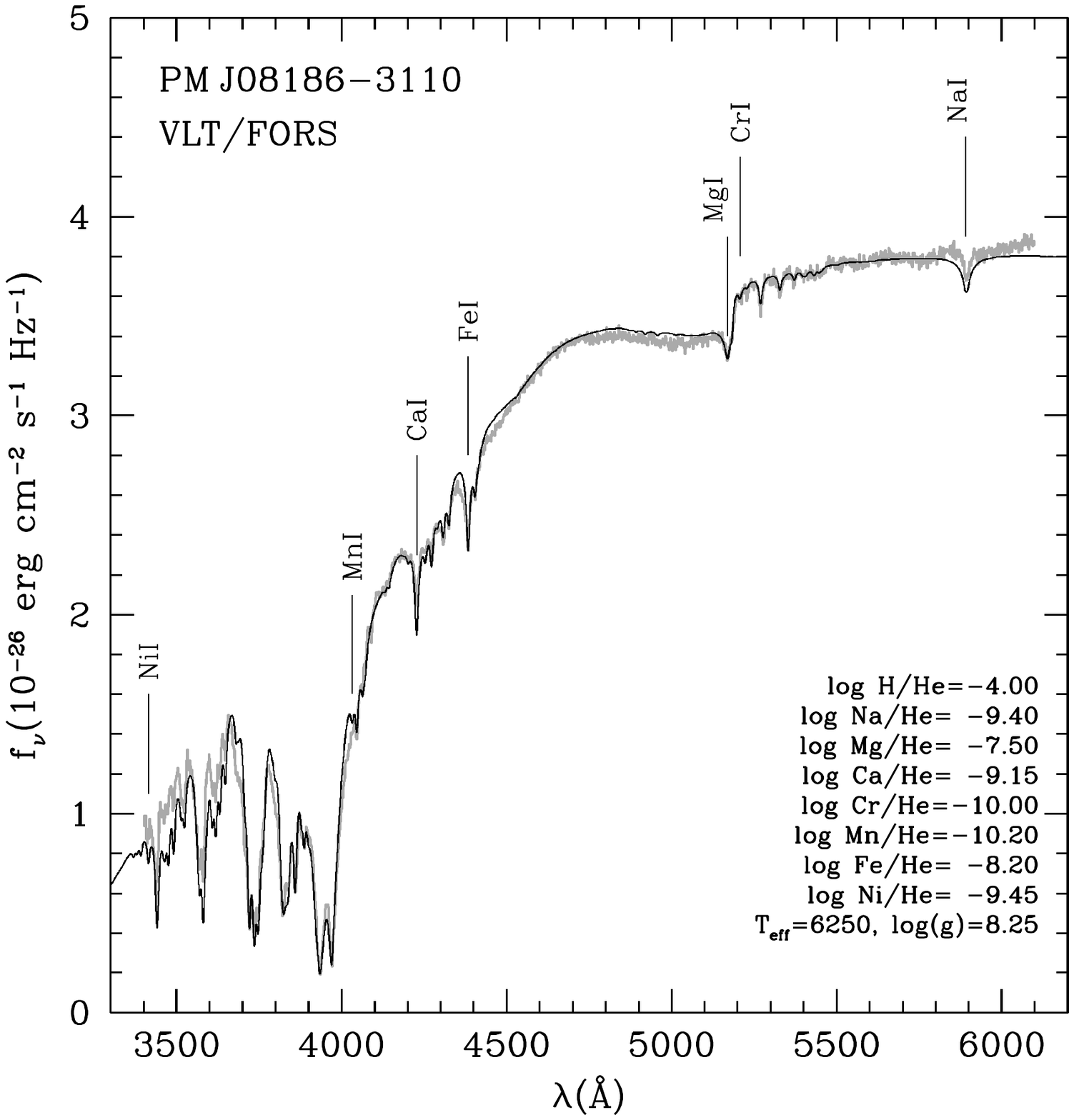}%
    \includegraphics[viewport=23 150 585 695,clip,width=0.9\columnwidth]{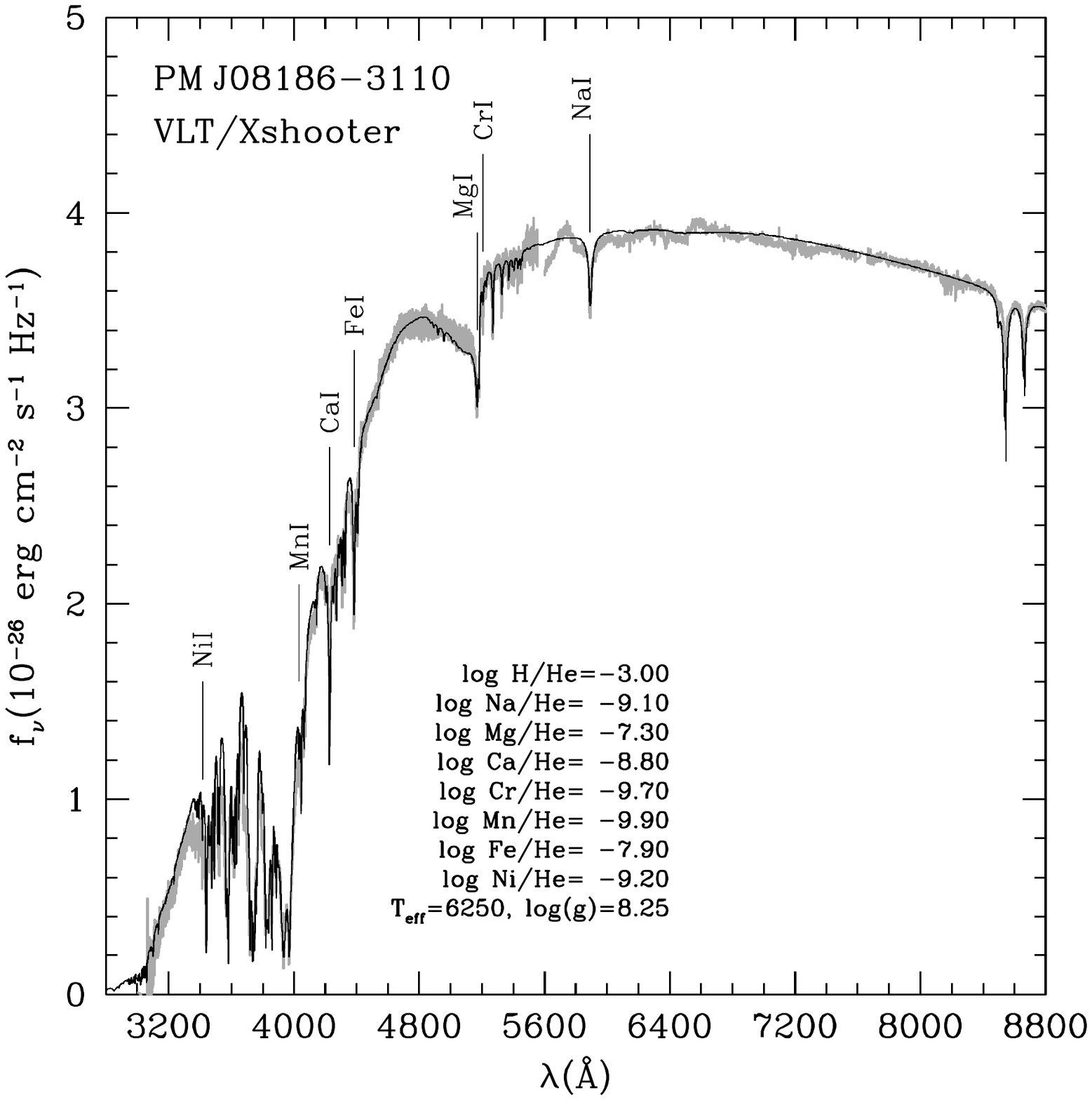}
    \caption{(Left:) FORS1 spectrum (grey) compared to the best fitting model spectrum (black) with tabulated model parameters. (Right:) Same as left panel but for the X-shooter spectrum (grey).}
    \label{fig_xshoot}
\end{figure*}

We have fitted both the VLT/FORS1 and VLT/X-shooter spectra to determine the atmospheric parameters. The CTIO/R-C spectrum appeared identical to the FORS1 spectrum and was not analyzed further. Fig.~\ref{fig_xshoot} shows the best-fitting model spectra compared to the observed FORS1 and X-shooter spectra and Table~\ref{tbl_param} summarises the parameters
of \pmj0818. We found that the best-fitting
model to the VLT/FORS1 and X-shooter spectra correspond to an effective temperature of 6250~K and surface gravity $\log{g} = 8.25$. However, the abundances in the X-shooter spectrum are systematically higher than those measured in the FORS1 spectrum by a factor of two, suggesting that the abundances vary over time. The atmosphere appears enriched in magnesium, $+0.63$ and $+0.54$ in the FORS1 and X-shooter spectra, respectively, and to a lesser extent calcium, $+0.22$ (FORS1) and $+0.28$ (X-shooter), relative to other elements. The presence of hydrogen is felt indirectly with a notable downward inflexion of the optical spectrum toward shorter wavelength. The hydrogen abundance measured in the X-shooter spectrum is a factor of ten larger than measured in the FORS1 spectrum. We noted a discrepancy between the best fit calcium abundance and the strength of the \ion{Ca}{ii} IR triplet suggesting that Lorentzian profiles are inadequate in this case, and require line profiles inferring both a unified theory of spectral line broadening
and accurate ab initio potential energies. The mass and cooling age were calculated using the evolutionary mass-radius relations of \citet{cam2017}. Although the mass and radius of \citet{cam2017} and \citet{ben1999} are consistent, the cooling age differs significantly due to additional mechanisms being
included in mass-radius relations of \citet{cam2017}. We estimate the errors on the temperature and surface gravity at 100~K and $0.05$ dex, respectively. Error bars on abundance measurements stem from the error bars on $T_{\rm eff}$ and $\log{g}$ are are estimated at 0.2~dex on average. Note that FORS1 and X-shooter abundance measurements vary in concert. 

The variable abundances are most likely due to the magnetic field distribution on the white dwarf surface which can result
in abundance spots. A variable magnetic field strength across the surface of the white dwarf is supported by changes in the 
longitudinal magnetic field noted in two spectropolarimetric measurements reported by \citet{bag2019}. Photometric
variability will most likely be strongest in the $u$ and $v$ bands because this is where metal lines are
strongest, although due to flux redistribution some variability should be present in the red and infrared spectral range as well. 
We searched for variability in the available $TESS$ photometric measurements but we did not find any significant 
variations. Using the 2 min cadence, we measured an upper limit on the semi-amplitude of flux variations of 1.8\%. Binning
the measurements to a 10 min cadence we reduced the limit on flux variations to 1.2\%. The best-fitting model
spectra from the FORS1 and X-shooter data predict a 1.4\% variation over the $TESS$ bandpass, which is consistent with the
measured upper limit. Photometric
variability has been detected in several magnetic white dwarfs. Rotating white dwarfs with very strong magnetic fields, magnetic
dichroism can explain the photometric observations, e.g., EUVE~J0317-85.5 \citep{ven2003}. In cool white dwarfs with weak 
magnetic fields, star spots have been proposed as the likely cause \citep{bri2013}. The variations in these white dwarfs are 
a few percent, comparable to our upper limit. Finally, \citet{kil2019} showed that the spectroscopic and photometric variations
observed in G183$-$35 were caused by a variable magnetic field and a chemically inhomogeneous surface 
composition rotating on a period of about 4 hr.

\begin{figure*}
\centering
\includegraphics[viewport=20 440 545 710,clip,width=0.9\textwidth]{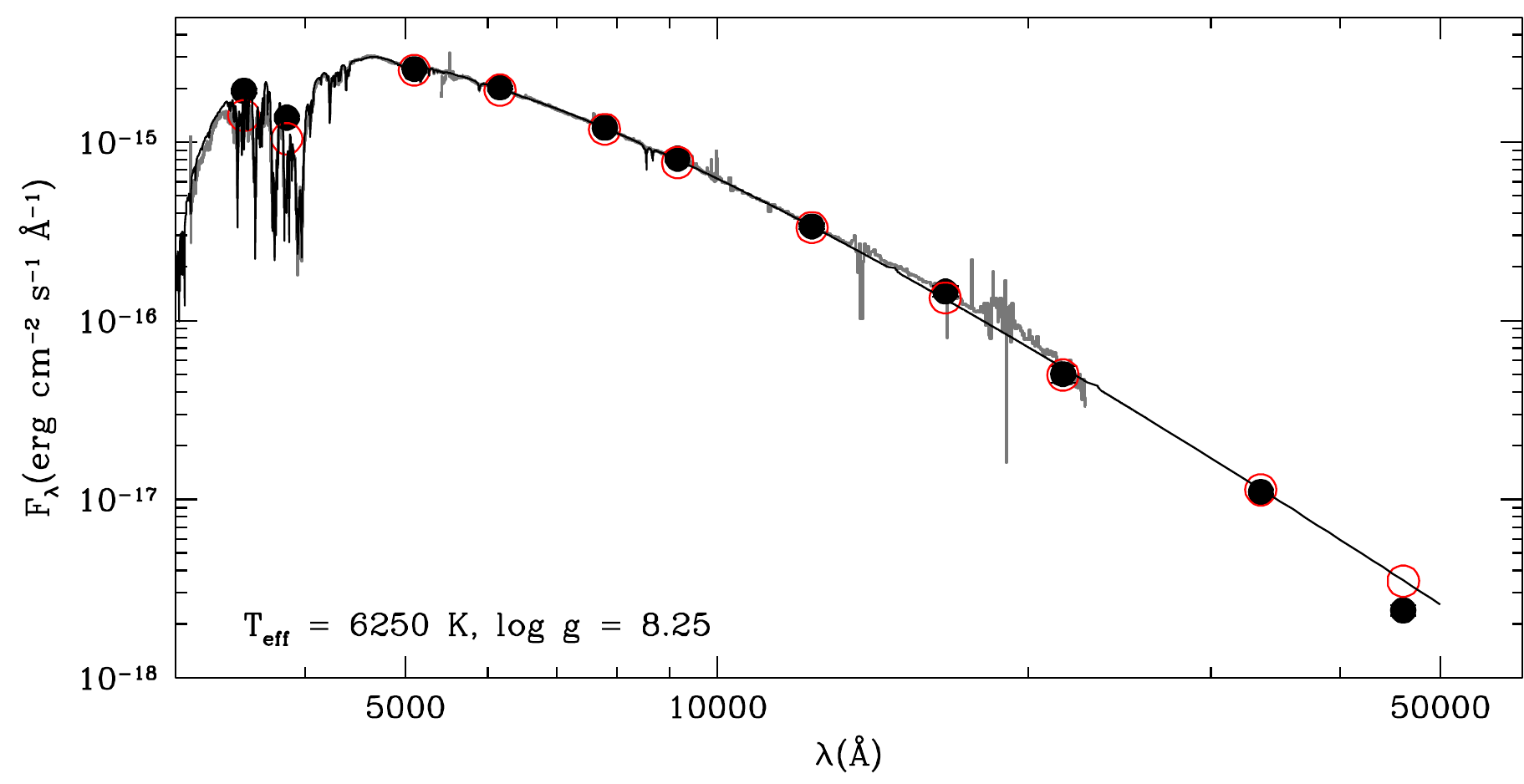}
\caption{The best fitting model spectrum compared to the X-shooter spectra (grey) and observed
photometric measurements (black dots). The model photometric points (red circles) are calculated from the
best fitting spectrum. The ultraviolet data points differ significantly from the X-shooter flux showing evidence of abundance variations.}
\label{fig_sed}
\end{figure*}

We compared the best-fitting model spectra to the observed photometric measurements. We used the X-shooter best-fitting 
spectrum to calculate model magnitudes. First we convolved the model spectrum with the SkyMapper \citep{bes2011}, 
2MASS \citep{skr2006} and WISE \citep{wri2010} bandpasses and then we calculated absolute magnitudes using the
evolutionary mass-radius relations of \citet{ben1999}. Fig.~\ref{fig_sed} shows the observed apparent magnitudes compared
to the model magnitudes which were converted from absolute magnitudes to apparent magnitudes using the Gaia distance. The comparison shows that the short wavelength bands, Skymapper $u$ and $v$, are systematically above the X-shooter flux level and show evidence of abundance variations.

\begin{table}
\caption{Properties of \pmj0818\ constrained by the Gaia distance.}
\label{tbl_const}
\centering
\begin{tabular}{cccc}
\hline
$T_{\rm eff}$ (K) & $\log{g}$ & Mass ($M_\odot$) & Radius ($R_\odot$) \\
\hline
  5800 & 8.02 & 0.572 &  0.01230 \\
  5900 & 8.07 & 0.606 &  0.01188 \\
  6000 & 8.13 & 0.639 &  0.01149 \\
  6100 & 8.18 & 0.672 &  0.01112 \\
  6200 & 8.22 & 0.705 &  0.01076 \\
  6300 & 8.27 & 0.736 &  0.01042 \\
  6400 & 8.31 & 0.767 &  0.01010 \\
  6500 & 8.36 & 0.796 &  0.00979 \\
  \hline
  \end{tabular}
\end{table}

Some atmospheric parameters of \pmj0818\ were published in the past and the effective temperatures from these studies are
systematically higher than what we have obtained, although individual measurements are consistent with our own within 1$\sigma$ error bars.
Using low resolution spectroscopy and assuming $\log{g}=8$, \citet{sub2008} measured an effective temperature
of $T_{\rm eff} = 6631\pm345$~K and $\log{\rm Ca/He} = -9.09\pm0.20$. A slightly lower effective temperature of
$T_{\rm eff} = 6463\pm370$~K, also assuming $\log{g}=8$, was determined by \citet{gia2012} along with $\log{\rm Ca/He} = -9.23$.
\citet{sub2017} measured a parallax of $\pi = 52.30\pm0.71$~mas and combined with $V, R, I$ photometric measurements and low
resolution spectra they constrained the parameters of \pmj0818\ to
$T_{\rm eff} = 6480\pm290$~K, $\log{g} = 8.22\pm0.07$ corresponding to a mass of $0.71\pm0.06\,M_\odot$. They also 
measured $\log{\rm Ca/He} = -9.26$. Using the Gaia parallax, $V, R, I$ photometry, and low resolution spectroscopy, 
\citet{cou2019} constrained the parameters of \pmj0818\ 
and obtained $T_{\rm eff} = 6535\pm645$~K, $\log{g} = 8.28\pm0.19$ and $\log{\rm Ca/He} = -8.99\pm0.04$, corresponding to a mass of $0.75\pm0.13\,M_\odot$.
The present analysis is more comprehensive than earlier ones as it benefited from a complete spectral coverage and a detailed abundance analysis.

\begin{table}
 \caption{Stellar and atmospheric parameters.}
 \label{tbl_param}
 \centering
 \begin{tabular}{lcc}
  \hline
  Parameter & \multicolumn{2}{c}{Measurement} \\
  \hline
            & FORS & X-shooter \\
  $T_{\rm eff}$ (K)   & \multicolumn{2}{c}{$6250$} \\
  $\log{g}$ (cgs)     & \multicolumn{2}{c}{$8.25$} \\
  Mass (M$_\odot$)    & \multicolumn{2}{c}{$0.72$} \\
  Distance (pc)       & \multicolumn{2}{c}{$19.4\pm0.01$} \\
  Age (Gyr)           & \multicolumn{2}{c}{4.2} \\
  $B_l$ (kG)          & $26\pm5$      & ... \\
  $B_S$ (kG)          &   ...         & $92\pm1$ \\
  $\log{\rm (H/He)}$  & $-4.00$  & $-3.0$ \\
  $\log{\rm (Na/He)}$ & $-9.40$ & $-9.10$ \\ 
  $\log{\rm (Mg/He)}$ & $-7.50$ & $-7.30$ \\
  $\log{\rm (Ca/He)}$ & $-9.15$ & $-8.80$ \\
  $\log{\rm (Cr/He)}$ & $-10.00$ & $-9.70$ \\
  $\log{\rm (Mn/He)}$ & $-10.20$ & $-9.90$ \\
  $\log{\rm (Fe/He)}$ & $-8.20$ & $-7.90$ \\
  $\log{\rm (Ni/He)}$ & $-9.45$ & $-9.20$ \\
  \hline
 \end{tabular}
\end{table}

\subsection{Magnetic field}

\begin{figure}
    \centering
    \includegraphics[viewport=35 150 585 710,clip,width=0.9\columnwidth]{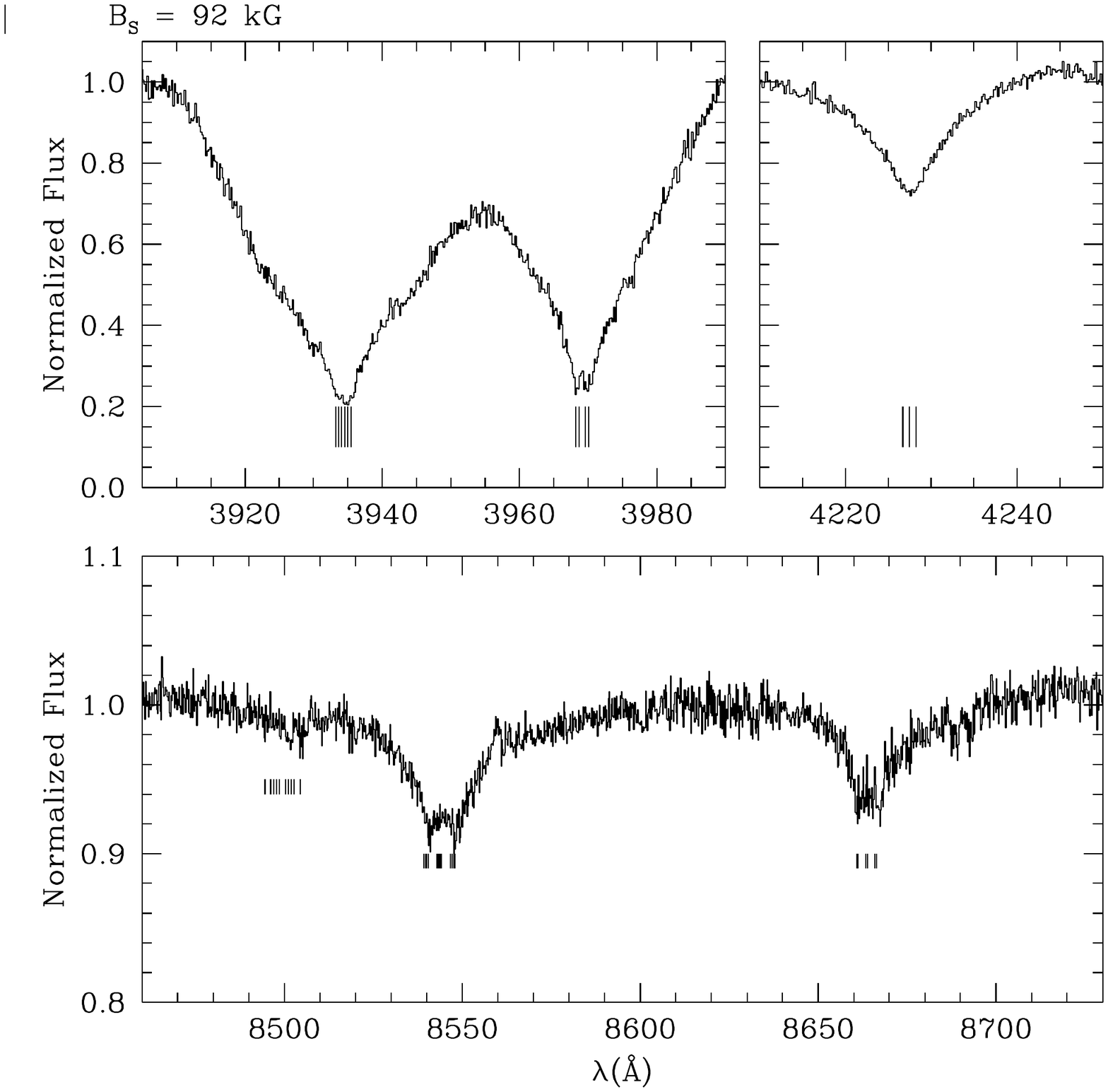}
    \caption{X-shooter spectra showing the Zeeman splitted or broadened \ion{Ca}{ii} and \ion{Ca}{i}
    lines. The tick marks indicate the predicted Zeeman splitting for magnetic field strength of 92~kG. The splitting is apparent in the strongest components of the \ion{Ca}{ii} IR triplet.}
    \label{fig_mag_ca}
\end{figure}

The average surface field strength ($B_S$) can be estimated from the splitted Zeeman components of spectral lines 
\citep[see][for details]{kaw2011}. We used the \ion{Ca}{ii} 8542 and 8662 \AA\ lines in the X-shooter spectrum to 
measure $B_S = 92\pm1$~kG. We measured a Barrycentric corrected radial velocity of $88.0$~km~s$^{-1}$. Fig.~\ref{fig_mag_ca} shows the Zeeman splitted or broadened \ion{Ca}{ii} and \ion{Ca}{i} lines observed in the
X-shooter spectrum. We checked the average surface field strength is consisted with other lines where the Zeeman splitting is less
pronounced but we clearly see a broadened core, these include \ion{Fe}{i} 4271.76 and \ion{Fe}{i} 5269.537 \AA.

\begin{figure}
    \centering
    \includegraphics[viewport=30 280 580 655,clip,width=0.9\columnwidth]{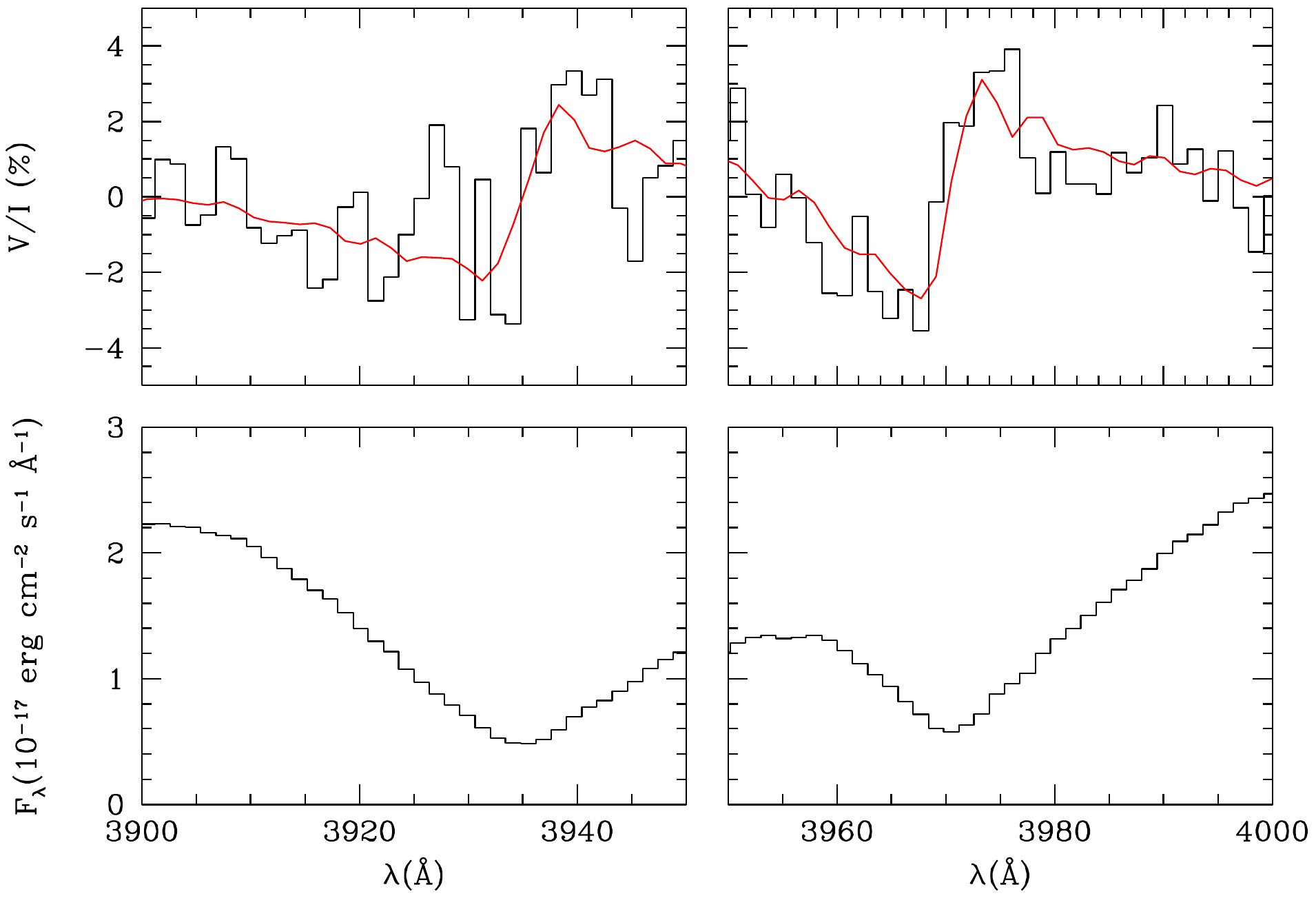}
    \caption{FORS1 circular polarization (top) and flux (bottom) spectra of the \ion{Ca}{ii} lines in \pmj0818.}
    \label{fig_fors1}
\end{figure}

We calculated the longitudinal magnetic field strength ($B_l$ in G) using the circularly polarised spectra obtained with FORS1 using
\begin{equation}
    B_l = \frac{v\,I}{4.67\times 10^{-13}\lambda^2 (dI/d\lambda)}
\end{equation}

where $v = V/I$ is the degree of circular polarization, $I$ is the total intensity, $\lambda$ is the wavelength in \AA, 
and $dI/d\lambda$ is the flux gradient. Fig.~\ref{fig_fors1} shows the circular polarization and flux spectra of 
the \ion{Ca}{ii} lines
with the best fitting $V/I$ spectrum for a longitudinal magnetic field strength of $B_l = 26\pm5$~kG. \citet{bag2019} 
measured a similar field of $20\pm2$~kG using the same set of observations. Note that the sign of circular polarization
in \citet{bag2019} is opposite to the sign used in the present paper. The second circular polarization spectrum obtained by \citet{bag2019} suggests that the magnetic field is variable supporting the likelihood of a varying abundance on the 
white dwarf surface.

\subsection{Kinematics}

We calculated the Galactic velocity components from the distance, proper motion and radial velocity of 
\pmj0818\ using \citet{joh1987}. We corrected for the Solar motion relative to the local standard of rest 
using ($U_\odot,V_\odot,W_\odot) = (11.10,12.24,7.25$)~km~s$^{-1}$ \citep{sch2010}. We measured a Barrycentric
corrected radial velocity of $v = 88.0\pm5.0$~km~s$^{-1}$ using the X-shooter spectrum. We subtracted a 
gravitational redshift of $V_{gr} = 43.3$~km~s$^{-1}$ to obtain a stellar velocity of $44.7$~km~s$^{-1}$. 
The Galactic velocity components of $(U,V,W) = (65,-54,-13)$~km~s$^{-1}$ place \pmj0818\ in the 
thick disc \citep{sou2003}. We also calculated the Galactic orbit properties using \textsc{galpy} 
\citep{bov2015} where we use the \texttt{MWPotential2014} Galactic potential. The eccentricity $e=0.33\pm0.02$
and the $z$-component of the angular momentum $L_Z = 1296$~kpc~km~s$^{-1}$ also places \pmj0818\ in the 
thick disc following \citet{pau2006}.
 
The age of the Galactic thick disc is about 10 Gyr \citep{sha2019} and therefore the total age of \pmj0818\ should be 
comparable. However, the cooling age of \pmj0818\ is 4.2 Gyr and assuming a thick disc metallicity, the mass and lifetime of the progenitor should be about $2.5\,M_\odot$ and 600 Myr \citep{rom2015}, respectively. Therefore, assuming single star evolution, the total age of \pmj0818\ should be significantly lower than 10 Gyr. 

The cooling age was calculated using the evolutionary tracks of \citet{cam2017} for hydrogen-deficient white dwarfs. They
show that in cool helium rich white dwarfs the high density atmosphere plays an important role in regulating their
cooling rates. Although \pmj0818\ is a cool helium-rich star, it is also polluted by heavy elements and the electrons
contributed by these elements, in particular iron, help increase the He$^-$ opacity and push the atmosphere to shallower depths than found in pure-helium atmospheres. This is also likely
to affect the cooling age estimate.

\section{Discussion}

The number of magnetic, polluted white dwarfs is growing and the incidence of magnetism was found to be higher in cool
polluted white dwarfs than in the general white dwarf population. For polluted hydrogen-rich (DAZ)
white dwarfs the incidence was found to be as high as 50\% for $T_{\rm eff} \lesssim 6000$~K \citep{kaw2014,kaw2019}.
For helium-rich white dwarfs this incidence is $21.6\pm3.3$\% \citep{hol2015,hol2017}. 

\subsection{Incidence of magnetism}

\begin{figure*}
    \centering
    \includegraphics[viewport=35 320 580 675, clip, width=0.8\textwidth]{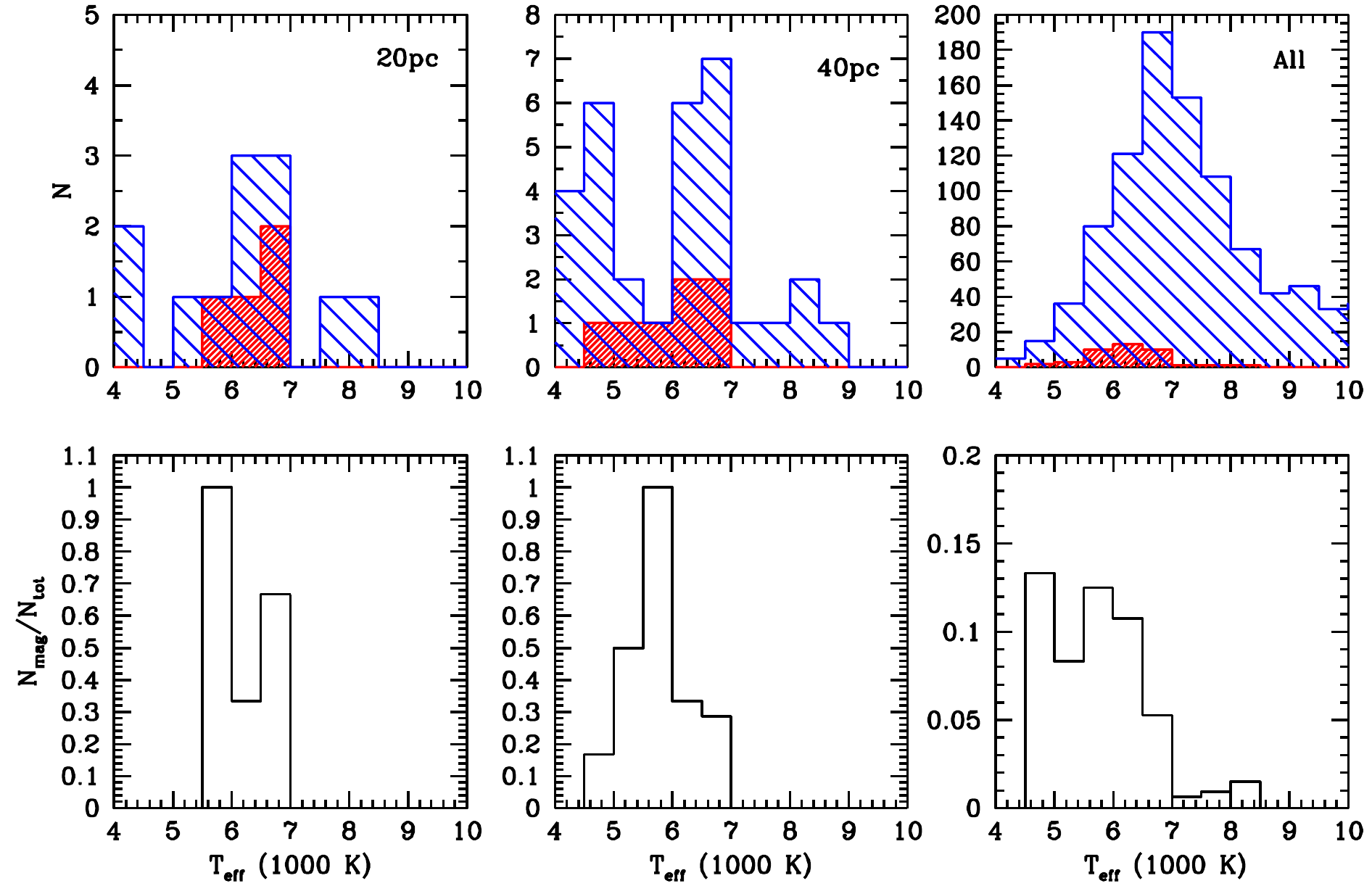}
    \caption{The incidence of magnetism in cool DZ white dwarf as a function of effective temperature.}
    \label{fig_incidence}
\end{figure*}

Recently, new magnetic DZ white dwarfs have been discovered in the Solar neighbourhood \citep[e.g.,][]{bag2019,tre2020}, 
and therefore their incidence can be revisited. We have assembled a sample of 
known DZ white dwarfs using the extensive lists of \citet{cou2019} and \citet{hol2017}. We also included new
DZ white dwarfs that lie within 40pc of the Sun \citep{tre2020}. We cross-correlated
this sample with the Gaia DR2 catalogue to obtain parallax measurements. These measurements allow us to build
volume limited samples.

There are 12 DZ white dwarfs within 20pc of the Sun, and of these 4 are magnetic. Fig.~\ref{fig_incidence} shows 
that the magnetic DZs are all cooler than 7000~K and the incidence of magnetism for DZs in this temperature 
range is 40\%. Extending the volume to 40pc, a total of 32 DZ white dwarfs are known, out of which 7 are magnetic.
Again, they are all below 7000~K, however the incidence of magnetism goes down to 27\% ($T_{\rm eff} < 7000$~K). 
Fig.~\ref{fig_incidence} also plots the incidence of magnetism for all known DZ white dwarfs and it shows that 
there are few magnetic DZs with a temperature between 7000 and 8500~K. Larger volume limited samples are needed
to confirm whether the trend of an increased incidence of magnetism with decreasing temperature is real.

\subsection{Abundance evolution}

The kinematics of \pmj0818 suggest that it is an old, thick disc star, and if we assumed
that the white dwarf atmosphere was polluted through the accretion or merging with
a planet or a low mass star, the abundance ratios of elements heavier than helium
would depend on the age, hence composition of the parent stellar system. \citet{spi2016} and \cite{bed2018} have
shown that the abundance of elements lighter than Fe should be higher in older stars and this should
be reflected in the composition of the accreted material onto polluted white dwarfs. Magnesium is 
enriched in \pmj0818 as compared to iron, chromium and manganese, which is predicted in old 
stellar systems.
However, diffusion alters the observed abundance ratios in white dwarf
atmospheres relative to the abundance pattern in the accretion 
sources. In this regard, \citet{hei2020} have shown that additional mixing criteria may
have a significant effect on the diffusion time scales of elements relative to each other. They also show that diffusion coefficients at the bottom of the deep convection zone of helium-rich white dwarfs may require considerable revision which prevents us from reaching firm conclusions on the evolution of abundance ratios as a function of the age of the source material. Moreover, the effect of  magnetic field on diffusion time scales remains largely unknown.

\section{Conclusions}

We have presented a spectroscopic analysis of the cool, He-rich, magnetic white dwarf \pmj0818, showing that it has an atmosphere enriched in magnesium relative to to sodium, calcium, chromium, manganese, iron and nickel. It is not know whether the magnesium dominance is a characteristic of an old, Fe-poor, parent environment, or it occurred later on during the accretion process. A population of Mg-rich polluted white dwarfs has already been identified and is representative of Earth-mantle material \citep{hol2018}. The presence of hydrogen in the atmosphere of \pmj0818 suggests that the accreted material, e.g., broken-up asteroids, contained ice. The recent detection of ice/water in the prototypical asteroid Ceres \citep{ray2020} supports this scenario. Signatures of accreted water in white dwarfs were detected in a handful of DZ white dwarfs \citep{far2013,gen2017}. This is the first time
an abundance pattern was measured for \pmj0818, with previous spectroscopic analyses resulted in the abundance
measurements of calcium, only. We have also shown that the abundance pattern changes between the FORS1 and X-shooter
spectra taken 10 yr apart, suggesting that \pmj0818 has abundance spots due to the distribution of the magnetic field on the
surface of the white dwarf. These abundance spots will reveal themselves as abundance variations over the rotational period, which
could be a few hours, such as was observed in G183-35 \citep{kil2019}. This hypothesis is supported by noted variations in the longitudinal field measured with FORS1 spectropolarimetry
and, independently, with WHT/ISIS spectropolarimetry \citep{bag2019} . The X-shooter spectrum allowed us to measure a weak surface average field of 92~kG. 

The kinematics of \pmj0818 suggests that it belongs to the thick disk, however, when using evolutionary models, the total age of 
\pmj0818 is significantly lower than the age of thick disc. This could mean that either \pmj0818 is not part of the thick disc but 
rather an outlier of the old thin disc population, or that additional factors or events could slow down the cooling rate of magnetic DZ white dwarfs. The third possibility could be that \pmj0818 is the result of a merger and therefore its cooling age was reset 
following the merging event.

We have shown that the incidence of magnetism in cool DZ white dwarfs is very high, similar to the incidence 
of magnetism in cool DAZ white dwarfs. Both populations are lacking counterparts in their warmer and younger counterparts
which suggests that the magnetic field formation and pollution of the atmosphere are linked to an accretion event that happens
late in the life of the white dwarf.

\section*{Acknowledgements}

 This study is based on observations made with ESO telescopes at the La Silla Paranal Observatory under programmes 080.D-0521 and 0100.D-0531. We thank Pierre Bergeron for a useful review of the paper. 

\section*{Data Availability}

The data underlying this article are available in the ESO Science Archive at http://archive.eso.org, and can be accessed under programmes
080.D-0521 and 0100.D-0531. 






\bsp	
\label{lastpage}
\end{document}